\documentclass{article}

\usepackage{amsthm}
\usepackage{amsmath}
\usepackage{natbib}
\usepackage[colorlinks,citecolor=blue,urlcolor=blue,filecolor=blue,backref=page]{hyperref}
\usepackage{graphicx}
\usepackage{bm}
\usepackage{amssymb}
\usepackage{mathabx}
\usepackage{algorithm}
\usepackage{algpseudocode}
\usepackage{fullpage}
\usepackage{setspace}
\begin{document}


\title{Minibatch Markov chain Monte Carlo Algorithms for Fitting Gaussian Processes}

\author{Matthew~J.~Heaton and Jacob~A.~Johnson}
\maketitle

\begin{abstract}
Gaussian processes (GPs) are a highly flexible, nonparametric statistical model that are commonly used to fit nonlinear relationships or account for correlation between observations.  However, the computational load of fitting a Gaussian process is $\mathcal{O}(n^3)$ making them infeasible for use on large datasets.  To make GPs more feasible for large datasets, this research focuses on the use of minibatching to estimate GP parameters.  Specifically, we outline both approximate and exact minibatch Markov chain Monte Carlo algorithms that substantially reduce the computation of fitting a GP by only considering small subsets of the data at a time.  We demonstrate and compare this methodology using various simulations and real datasets.
\end{abstract}

\section{Introduction}\label{intro}

\subsection{Problem Background}
Let $Y(\bm{s})$ be a response variable measured at location $\bm{s} \in \mathcal{D} \subset \mathbb{R}^d$. $Y(\bm{s})$ is said to follow a Gaussian process (GP) if for any finite collection of locations $\bm{s}_1,\dots,\bm{s}_n$ then
\begin{align}
    \bm{Y} &\sim \mathcal{N}(\bm{\mu}, \bm{\Sigma})\label{fullModel}
\end{align}
where $\bm{Y} = (Y(\bm{s}_1),\dots,Y(\bm{s}_n))'$ and $\mathcal{N}(\bm{\mu}, \bm{\Sigma})$ is the multivariate normal distribution with mean vector $\bm{\mu} = (\mu(\bm{s}_1),\dots,\mu(\bm{s}_n))'$ and covariance matrix $\bm{\Sigma} = \{\sigma_{ij}\}_{i,j=1}^n$.  In Gaussian processes, the mean vector is typically taken to be $\mu(\bm{s}) = \bm{x}'(\bm{s})\bm{\beta}$ where $\bm{x}(\bm{s}) = (x_0(\bm{s}),\dots,x_P(\bm{s}))'$ is a vector of covariates and $\bm{\beta} = (\beta_0,\dots,\beta_P)'$ is a vector of linear coefficients and, most commonly, $x_0(\bm{s}) = 1$ for all $\bm{s} \in \mathcal{D}$ so that $\beta_0$ corresponds to an intercept term.  In contrast, the covariance is governed by a covariance function $K(\cdot)$ (also commonly referred to as a kernel) such that
\begin{align}
    \sigma_{ij} = K(\bm{s}_i, \bm{s}_j \mid \bm{\phi})
\end{align}
where $\bm{\phi} = (\phi_1,\dots,\phi_Q)'$ is a vector of parameters underlying the covariance (typically consisting of range and smoothness parameters).  In the statistics literature, the most common family of covariance functions is the Mat\'{e}rn family which includes both the exponential and Gaussian covariance as special cases (see \citealt{matern1960spatial} and \citealt{genton2001classes} for details on covariance functions).

The power and flexibility of the GP is well documented across numerous papers and books \citep[see][for examples]{banerjee2003hierarchical, gelfand2010handbook, cressie2015statistics, gelfand2016spatial, wikle2019spatio}.  However, the Gaussian process has been limited in more recent years due to the computational complexity associated with model fitting.  Specifically, if $\bm{X}$ is the $n \times (P+1)$ matrix of linear covariates, estimates for the parameters $\bm{\Theta} = (\bm{\beta}',\bm{\phi}')'$ can be obtained by either maximizing the likelihood,
\begin{align}
    \mathcal{L}(\bm{\Theta}) &\propto |\bm{\Sigma}|^{-1/2}\exp\left\{-\frac{1}{2}(\bm{Y}-\bm{X\beta})'\bm{\Sigma}^{-1/2}(\bm{Y}-\bm{X\beta})\right\}
    \label{fullLike}
\end{align}
or using this likelihood in conjunction with a prior for $\bm{\Theta}$ to obtain the corresponding posterior distribution in a Bayesian framework.  The likelihood in \eqref{fullLike} immediately shows an issue with using the GP.  Specifically, the necessity of storing and calculating the inverse and determinant of a $n\times n$ matrix is prohibitively expensive.

Given the computational complexities mentioned above, much research regarding GPs has focused on how to apply them to large datasets.  Early solutions leaned on the Karhunen-Lo\`{e}ve theorem and proposed low rank approximations \citep[see][]{banerjee2008gaussian, cressie2008fixed, sang2012full} using carefully constructed basis functions.  Simultanesouly, other groups investigated the use of compactly-supported covariance functions \citep{furrer2006covariance, kaufman2008covariance, noack2023exact} or partitioning \citep{konomi2014adaptive, heaton2017nonstationary} to introduce sparsity into the covariance matrix to ease the computational burden of matrix inversion.  The limitations of these early solutions quickly became apparent \citep{stein2014limitations} so that research in this area shifted to where it predominately resides today by using either sparse precision matrices often constructed using Vecchia (or nearest-neighbor) approximations \citep{datta2016hierarchical, datta2016cholesky, katzfuss2021general} or large computing clusters \citep{abdulah2018exageostat} or both.  Reviews on these methods and their comparative performances on datasets of various size and complexity are available in \citet{heaton2019case}, \citet{huang2021competition} and \citet{abdulah2022second}.

The majority of the above methods were developed primarily in the statistics community as approximations to a full Gaussian process.  In contrast, the computer science community has taken a different approach to computational scalability based on minibatch sampling of the dataset to perform inference.  In terms of maximum likelihood, the stochastic gradient descent algorithm and its variants \citep{bottou2012stochastic} has dominated the computer science literature for optimization.  In the Bayesian context, \citet{korattikara2014austerity} proposed a sequential hypothesis test for Metropolis-Hastings (MH) proposals based on a fraction of the full dataset.  Building on this seminal work, other minibatch MH algorithms were then developed by \citet{seita2016efficient} and \citet{bardenet2014towards}.  In more recent years, minibatched approaches in Markov chain Monte Carlo (MCMC) has grown to include tempered methods \citep{li2017mini}, Gibbs sampling \citep{de2018minibatch} and gradient-based proposals \citep{wu2022mini} with a review provided in \citet{bardenet2017markov}.  However, recent work by \citet{johndrow2020no} has warned that these minibatch approaches don't necessarily equate to improved sampling from the posterior distribution.

\subsection{Research Goals and Contributions}
The issue with the above minibatch solutions is that none of them are developed in the context of GPs.  Notably, the likelihood in \eqref{fullLike} requires the full dataset to compute and is not represented as a product of independent likelihoods as is needed for each of the minibatch algorithms above.  As such, the purpose of this research is to merge the statistical and computer science approaches by developing approaches to use minibatching in Markov chain Monte Carlo (MCMC) algorithms to fit GPs to large datasets within the Bayesian paradigm.  While \citet{saha2023incorporating} have also considered subsampling, our approach here is inherently different.  That is, \citet{saha2023incorporating} explicitly model a subsampling mechanism while we choose to use subsampling to either approximate a Metropolis-Hastings acceptance probability or the parameters of the complete conditional distribution (when known).

Our key to using minibatches for fitting GPs is to first represent the full likelihood in \eqref{fullLike} as a series of conditional distributions using the Vecchia approximation \citep[see][]{katzfuss2021general}.  Under this Vecchia approximation, \eqref{fullLike} can be written as a product of conditional probability density functions which is then ported into minibatch approaches such as those cited above.  In the case of the GP presented above, however, we note that the complete conditional distribution of some of the parameters in \eqref{fullLike} are conjugate under certain prior specifications (for example, the linear regression coefficients $\bm{\beta}$ are conjugate under a Gaussian prior).  Using these conjugate forms can greatly increase the efficiency of any MCMC algorithm. Hence, while we develop minibatch updating schemes for non-conjugate parameters, we also exploit the known form of conjugate complete conditional distributions by using appropriate minibatched approximations of the complete conditional distribution for conjugate model parameters.

The remainder of this paper is outlined as follows.  Section \ref{chap2} provides details of how to use minibatching in a MCMC algorithm.  Section \ref{examples} evaluates the upsides and downsides of using minibatching on simulated and real datasets.  Finally, Section \ref{conc} provides discussion and areas of future research.

\section{Methods}\label{chap2}
This section describes the details of how we use minibatching within an MCMC algorithm for GPs. 
Specifically, Section \ref{prelim} sets up the GP model including the Vecchia approximation framework and a MCMC algorithm using all available data.  Section Section \ref{minicc} discusses a minibatch updating scheme for conjugate parameters and \ref{barker} discusses options for an accept-reject rule for non-conjugate parameters based on minibatches.  Finally, Section \ref{implement} describes some nuances and details of implementing the minibatch MCMC algorithms.

\subsection{Preliminaries}\label{prelim}
As in the previous section, let $\bm{Y} = (Y(\bm{s}_1),\dots,Y(\bm{s}_n))'$ be a vector of response variables measured at the finite set of locations $\bm{s}_1,\dots,\bm{s}_n$ and $\bm{X}$ be a $n \times (P+1)$ matrix of covariates that are linearly related to $\bm{Y}$.  For purposes of this research, we let $Y(\bm{s})$ follow a Gaussian process such that the $ij^{th}$ entry of the covariance matrix $\bm{\Sigma}$ is given by
\begin{align}
    K(\bm{s}_i, \bm{s}_j \mid \sigma^2, \omega, \bm{\phi}) &= \begin{cases} \sigma^2 &\text{if } i=j \\
    \sigma^2(1-\omega)\rho(\bm{s}_i, \bm{s}_j \mid \bm{\phi}) & \text{if } i \neq j
    \end{cases}\label{covFX}
\end{align}
where $\sigma^2$ is the total variance (also referred to as the sill in spatial statistics terminology), $\omega\sigma^2$ for $\omega \in [0,1]$ is the nugget effect, $(1-\omega)\sigma^2$ is the partial sill and $\rho(\bm{s}_i, \bm{s}_j)$ is a positive definite correlation function (e.g.\ Mat\'{e}rn, Exponential, Gaussian, etc.) parameterized by $\bm{\phi}$.  Under this choice of covariance function, the covariance matrix takes the simple form $\bm{\Sigma} = \sigma^2\bm{R} = \sigma^2(\omega\bm{I}+(1-\omega)\bm{M})$ where $\bm{M} = \{\rho(\bm{s}_i,\bm{s}_j\mid \bm{\phi})\}_{i,j=1}^n$.

Because the joint distribution of $\bm{Y}$ is multivariate Gaussian, the likelihood for $\bm{\Theta} = (\bm{\beta}, \sigma^2, \omega, \bm{\phi}')'$ in \eqref{fullLike} can be written as a series of conditional distributions such that:
\begin{align}
\mathcal{L}(\bm{\Theta}) &= f_1(Y(\bm{s}_1) \mid \bm{\Theta})\prod_{i=2}^n f_i(Y(\bm{s}_i) \mid Y(\bm{s}_1),\dots,Y(\bm{s}_{i-1}),\bm{\Theta})
\label{condllike}
\end{align}
where each $f_i(\cdot)$ is a univariate Gaussian probability density function (PDF) with mean $\mu_i$ and variance $\sigma^2 v_i$.  Specifically, the mean terms are given by:
\begin{align}
    \mu_i &= \begin{cases} \bm{x}'(\bm{s}_1)\bm{\beta} &\text{if } i = 1 \\
    \bm{x}'(\bm{s}_i)\bm{\beta} + \bm{R}(i, \mathcal{N}_i)\bm{R}^{-1}(\mathcal{N}_i, \mathcal{N}_i)(\bm{Y}_{\mathcal{N}_i}-\bm{X}_{\mathcal{N}_i}\bm{\beta}) & \text{if } i > 1
    \end{cases}
    \label{mui}
\end{align}
where $\mathcal{N}_i = \{1,\dots,i-1\}$ is the set of points preceding observation $i$, $\bm{Y}_{\mathcal{N}_i}$ is the set of $Y(\bm{s})$ corresponding to $\mathcal{N}_i$, $\bm{X}_{\mathcal{N}_i}$ are the rows of $\bm{X}$ corresponding to $\mathcal{N}_i$ and $\bm{R}(\mathcal{A},\mathcal{B})$ is the corresponding $\#\mathcal{A} \times \#\mathcal{B}$ correlation matrix from the above correlation function where $\#$ denotes cardinality.  Likewise, we define
\begin{align}
 v_i &= \begin{cases} 1 & \text{if } i = 1 \\
 1- \bm{R}(i, \mathcal{N}_i)\bm{R}^{-1}(\mathcal{N}_i, \mathcal{N}_i)\bm{R}(\mathcal{N}_i,i) & \text{if } i > 1
 \end{cases}
 \label{vi}
\end{align}

Importantly, factoring the full likelihood from \eqref{fullLike} using a series of conditional distributions as in \eqref{condllike} does not circumvent the computational difficulties associated with GPs.  Specifically, the forms for $\mu_i$ and $v_i$ in \eqref{mui} and \eqref{vi} still require dealing with large matrices through $\bm{R}(\mathcal{N}_i, \mathcal{N}_i)$ because $\#\mathcal{N}_i$ grows as $i \rightarrow n$.  Hence, we adopt the Vecchia process approximation framework \citep[see][]{datta2016hierarchical, datta2016cholesky, katzfuss2021general} by redefining $\mathcal{N}_i$ be the set of the $M$ nearest neighbors of $\bm{s}_i$ in terms of Euclidean distance.  In this way, $\#\mathcal{N}_i$ is at most $M$ so that $\bm{R}(\mathcal{N}_i, \mathcal{N}_i)$ is also at most $M \times M$ and can be dealt with computationally.  This Vecchia approximation relies on the assumption that all the information about $Y(\bm{s}_i)$ in the conditional distribution $f_i(Y(\bm{s}_i))$ can be adequately summarized by the $M$ nearest neighbors to $Y(\bm{s}_i)$ among $Y(\bm{s}_1),\dots,Y(\bm{s}_{i-1})$.  We note that \citet{katzfuss2021general} discuss the impact of observation ordering on this assumption and recommend certain orderings of the observations to obtain better approximations.  For purposes of this research, we assume that our observations have already been ordered according to these suggestions.

The unknown parameters of our Gaussian process model are the linear coefficients $\bm{\beta}$, the sill $\sigma^2$ and the correlation parameters which include the nugget term $\omega \in [0,1]$ and any correlation function parameters $\bm{\phi}$.  Our focus here is on a Bayesian estimation paradigm for these parameters but we note that maximum likelihood can also be used to obtain estimates.  Generally, a MCMC algorithm for sampling from the posterior distribution of these parameters can be done via Gibbs sampling where at each iteration, as we show below, $\bm{\beta}$ and $\sigma^2$ can be directly drawn from their complete conditional distributions while indirect sampling (e.g.\ Metropolis or Metropolis-Hastings) is used to draw $\bm{\theta} = (\omega, \bm{\phi}')'$.  Details are as follows.

In the Bayesian paradigm, we \textit{a priori} assume $\beta_p \overset{iid}{\sim} \mathcal{N}(m_p, s^2_p)$ for $p=0,\dots,P$ and $\sigma^2 \sim \mathcal{IG}(a_\sigma, b_\sigma)$ where $\mathcal{IG}$ denotes the inverse-gamma distribution with shape $a_\sigma$ and rate $b_\sigma$.  We specifically choose these priors because the associated complete conditional distribution of each $\beta_p$ can be shown to be conjugate with respect to the Vecchia likelihood in \eqref{condllike}.  Specifically, through some algebraic manipulation, the complete conditional distributions for $\beta_p$ and $\sigma^2$ are given by
\begin{align}
    \beta_p \mid - &\sim \mathcal{N}\left( \left[ \frac{\sum_{i=1}^nq_1(\bm{s}_i)}{\sigma^2} + \frac{1}{s^2_p} \right]^{-1}\left( \frac{\sum_{i=1}^nq_2(\bm{s}_i)}{\sigma^2} + \frac{m_p}{s^2_p}  \right), \left[ \frac{\sum_{i=1}^nq_1(\bm{s}_i)}{\sigma^2} + \frac{1}{s^2_p} \right]^{-1} \right)
    \label{betaCC} \\
    \sigma^2 \mid - &\sim \mathcal{IG}\left( \frac{n}{2} + a_\sigma, \frac{\sum_{i=1}^nq_3(\bm{s}_i)}{2} + b_\sigma \right) \label{sig2CC}
\end{align}
where ``$-$'' denotes all other parameters \textit{and} the data and the quantities $q_1(\bm{s}_i), q_2(\bm{s}_i)$ and $q_3(\bm{s}_i)$ are given by
\begin{align}
    q_1(\bm{s}_i) &= \frac{1}{v_i}\left[x_p(\bm{s}_i) - \bm{R}(i, \mathcal{N}_i)\bm{R}^{-1}(\mathcal{N}_i, \mathcal{N}_i)\bm{X}_{\mathcal{N}_i,p}\right]^2 \\
    q_2(\bm{s}_i) &= \frac{1}{v_i}\left[x_p(\bm{s}_i) - \bm{R}(i, \mathcal{N}_i)\bm{R}^{-1}(\mathcal{N}_i, \mathcal{N}_i)\bm{X}_{\mathcal{N}_i,p}\right]r_p(\bm{s}_i) \\
    q_3(\bm{s}_i) &= \frac{1}{v_i}\left[Y(\bm{s}_i) - \mu_i)\right]^2
\end{align}
where $\bm{X}_{\mathcal{N}_i,p}$ is the $p^{th}$ column of $\bm{X}_{\mathcal{N}_i}$, 
\begin{align*}
    r_p(\bm{s}_i) = Y(\bm{s}_i) - \bm{x}'_{-p}(\bm{s}_i)\bm{\beta}_{-p} - \bm{R}(i, \mathcal{N}_i)\bm{R}^{-1}(\mathcal{N}_i, \mathcal{N}_i)(\bm{Y}_{\mathcal{N}_i}-\bm{X}_{\mathcal{N}_i,-p}\bm{\beta}_{-p})
\end{align*}
$\bm{X}_{\mathcal{N}_i,-p}$ is all columns of $\bm{X}_{\mathcal{N}_i}$ \textit{except} the $p^{th}$ column and $\bm{\beta}_{-p}$ are all $\beta$ coefficients \textit{except} $\beta_p$.  We note that the full $\bm{\beta}$ vector is conjugate as a multivariate normal but for purposes of exposition we work with each $\beta_p$ individually but the algorithms below can be implemented for the full $\bm{\beta}$ vector.

Unlike $\bm{\beta}$ and $\sigma^2$, there are no conjugate priors for $\omega$ and $\bm{\phi}$.  Further, because such priors may change depending on which correlation function is chosen, for purposes of this research, we group them together and generally assume $\bm{\theta} = (\omega, \bm{\phi}')' \sim \pi(\cdot)$ for some parametric prior $\pi(\cdot)$.  Due to lack of conjugacy, MCMC simulation of $\bm{\theta}$ is done via accept-reject style algorithms.  For purposes of the minibatch algorithms below, we write the acceptance rule for a proposal $\bm{\theta}_{\text{prop}}$ given the current draw $\bm{\theta}_{\text{cur}}$ as
\begin{align}
    \Delta(\bm{\theta}_{\text{prop}}, \bm{\theta}_{\text{cur}}) &= \sum_{i=1}^n \Lambda_i - \log\left(\frac{\pi(\bm{\theta}_{\text{prop}})g(\bm{\theta}_{\text{cur}} \mid \bm{\theta}_{\text{prop}})}{\pi(\bm{\theta}_{\text{cur}})g(\bm{\theta}_{\text{prop}} \mid \bm{\theta}_{\text{cur}})}\right) + L
    \label{accept}
\end{align}
where $\Lambda_i = \log[f_i(Y(\bm{s}_i) \mid \bm{Y}_{\mathcal{N}_i}, \bm{\beta}, \sigma^2, \bm{\theta}_{\text{prop}}) / f_i(Y(\bm{s}_i) \mid \bm{Y}_{\mathcal{N}_i}, \bm{\beta}, \sigma^2, \bm{\theta}_{\text{cur}})]$ is the log-likelihood ratio for observation $i$, $g(\cdot)$ is the proposal distribution and $L$ is a random variable (note in Metropolis-Hastings algorithms $L = -\log(U)$ where $U \sim \mathcal{U}(0,1)$).  Accepting a proposed $\bm{\theta}_{\text{prop}}$ occurs if and only if $\Delta(\bm{\theta}_{\text{prop}}, \bm{\theta}_{\text{cur}}) > 0$.

\subsection{Minibatch Approximation of Quantities in Complete Conditional Distributions}\label{minicc}
The complete conditional distributions of $\bm{\beta}$ and $\sigma^2$ in \eqref{betaCC} and \eqref{sig2CC}, respectively, are also computationally challenging due to the large summations of $q_1(\bm{s}_i)$, $q_2(\bm{s}_i)$ and $q_3(\bm{s}_i)$. For conjugate parameters, a minibatch approximation of the summations is given by
\begin{align}
    \sum_{i=1}^n q_j(\bm{s}_i) = n\widebar{q}_{j,n} &= n\frac{1}{n}\sum_{i=1}^n q_j(\bm{s}_i) \notag \\
    &\approx n\widebar{q}_{j, B} = n\frac{1}{B}\sum_{i \in \mathcal{B}} q_j(\bm{s}_i) \label{miniCC}
\end{align}
for $j \in \{1, 2, 3\}$ where $\mathcal{B} \subseteq \{1,\dots,n\}$ is a minibatch with $\#\mathcal{B} = B$.  

Because $\widebar{q}_{j, B} \neq \widebar{q}_{j,n}$, we only want to replace $n\widebar{q}_{j,n}$ with $n\widebar{q}_{j,B}$ if they are sufficiently close.  Note that by the central limit theorem,
\begin{align}
    n\widebar{q}_{j,B} &\overset{d}{\rightarrow} \mathcal{N}\left(n\widebar{q}_{j,n}, \frac{n^2}{B}\sqrt{\frac{n - B}{n-1}}\sigma^2_{q_j}\right)
    \label{qDist}
\end{align}
where $\sigma^2_{q_j} = \mathbb{V}\text{ar}(q_j(\bm{s}_i))$ and, $\sqrt{(n-B)/(n-1)}$ is the finite population correction factor which ensures that $n\widebar{q}_{j, B} = n\widebar{q}_{j,n}$ when $B = n$.  Under this distribution, $n\widebar{q}_{j, B} \approx n\widebar{q}_{j,n}$ when $(n^2/B)\sqrt{(n - B)/(n-1)}\sigma^2_{q_j}$ is small which occurs as $B \rightarrow n$ suggesting that larger minibatch sizes should be preferred when the computation is reasonable.


\subsection{Minibatch Acceptance Test for Non-conjugate Parameters}\label{barker}
Similar to the complete conditional distributions, when $n$ is large, the sum in \eqref{accept} is slow computationally.  Hence, we use the minibatch approximation
\begin{align}
    \sum_{i=1}^n \Lambda_i = n\widebar{\Lambda}_n &= n\frac{1}{n}\sum_{i=1}^n \Lambda_i \notag \\
    &\approx n\widebar{\Lambda}_B = n\frac{1}{B}\sum_{i \in \mathcal{B}} \Lambda_i \notag
\end{align}
where $\mathcal{B} \subseteq \{1,\dots,n\}$ is a random minibatch of the observations and $\#\mathcal{B} = B$.  Because our minibatch approximation is an average, by the central limit theorem 
\begin{align}
    n\widebar{\Lambda}_B \overset{d}{\rightarrow} \mathcal{N}\left(n\bar{\Lambda}_n, \frac{n^2}{B}\sqrt{\frac{n - B}{n - 1}}\sigma_{\Lambda}^2\right) \label{lbarDist}
\end{align}
where $\sigma^2_\Lambda = \mathbb{V}\text{ar}(\Lambda_i)$ and $\sqrt{(n-B)/(n-1)}$ is the finite population correction factor which ensures that $\widebar{\Lambda}_B = \widebar{\Lambda}_n$ when $B = n$.

One possible approach to using minibatches in accept-reject type algorithms is then to simply replace $\bar{\Lambda}_n$ in \eqref{accept} with $\bar{\Lambda}_B$ and set $L = -\log(U)$ in a minibatch approximation of the Metropolis-Hastings algorithm.  This approach has the advantage that a fixed batch can be used.  In this way, we can randomly determine minibatches prior to running the MCMC algorithm and save on computation time.  However, because $\bar{\Lambda}_n \neq \bar{\Lambda}_B$, we detail an alternative minibatch approach that accounts for this discrepancy.

\begin{figure}[tb]
    \centering
    \includegraphics[scale=.6]{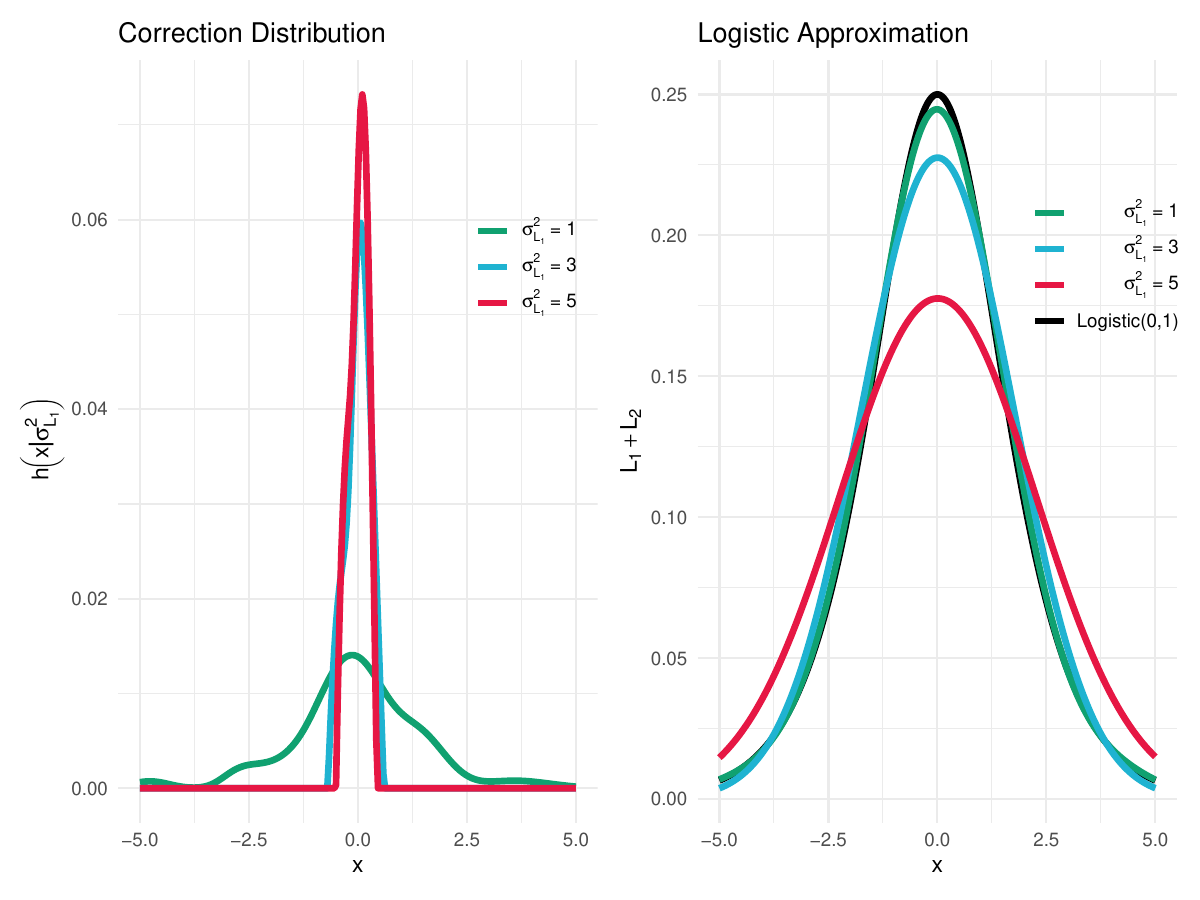}
    \caption{(Left) The correction distribution $h(\cdot)$ along with the resulting approximation (right) to the standard logistic distribution from convolution with a Gaussian distribution with variance $\sigma^2_{L_1}$.  Note that the approximation increases in accuracy for smaller $\sigma^2_{L_1}$.}
    \label{fig:cdist}
\end{figure}

To rewrite \eqref{accept} in terms of $\widebar{\Lambda}_B$, we can follow the approach of \citet{seita2016efficient}.  Specifically, as shown in \citet{barker1965monte}, if $L$ follows a standard logistic distribution then the acceptance test $\Delta(\bm{\theta}_{\text{prop}}, \bm{\theta}_{\text{cur}}) > 0$ in \eqref{accept} maintains detailed balance.  As such, let $L$ follow a standard logistic distribution and be represented as the sum $L = L_1 + L_2$ where $L_1 \sim \mathcal{N}(0, \sigma^2_{L_1})$ and $L_2 \sim h(\cdot)$ where \citet{seita2016efficient} refers to  $h(\cdot)$ as a ``correction'' distribution of the Gaussian distribution to the standard logistic random variable. By convolution, $\ell(z) \approx \sum_{x \in \mathcal{X}} f_{\text{Gaus}}(z-x \mid 0, \sigma^2_{L_1}) h(x \mid \sigma^2_{L_1})$ where $\ell(z)$ is the standard logistic PDF and $\mathcal{X}$ is some fine grid on the support of the standard logistic distribution.  Under this convolution construction, for a given $\sigma^2_{L_1}$, $h(x \mid \sigma^2_{L_1})$ can be estimated via penalized least squares (e.g.\ LASSO) with positivity constraints.  Figure \ref{fig:cdist} displays the construction of a standard logistic via convolution in this manner.  Specifically, the left panel of Figure \ref{fig:cdist} displays the correction distribution while the right panel displays the PDF from the convolution of this correction distribution with a Gaussian distribution for a given $\sigma^2_{L_1}$.  Notably, as can be seen in Figure \ref{fig:cdist}, this representation of $L = L_1 + L_2$ is best when, approximately, $\sigma^2_{L_1} \leq 3$ but grows in accuracy as $\sigma^2_{L_1}$ decreases.  

Under $L = L_1 + L_2$ as above, if we set $\sigma^2_{L_1} = (n^2/B)\sqrt{(n - B)/(n - 1)}\sigma_{\Lambda}^2$ then $n\widebar{\Lambda}_B = n\widebar{\Lambda}_n + L_1$ so that the acceptance rule in \eqref{accept} can be rewritten as
\begin{align}
    \Delta(\bm{\theta}_{\text{prop}}, \bm{\theta}_{\text{cur}}) &= n\widebar{\Lambda}_n + \log\left(\frac{\pi(\bm{\theta}_{\text{prop}})g(\bm{\theta}_{\text{cur}} \mid \bm{\theta}_{\text{prop}})}{\pi(\bm{\theta}_{\text{cur}})g(\bm{\theta}_{\text{prop}} \mid \bm{\theta}_{\text{cur}})}\right) + L_1 + L_2 \notag \\
    &= n\widebar{\Lambda}_B + \log\left(\frac{\pi(\bm{\theta}_{\text{prop}})g(\bm{\theta}_{\text{cur}} \mid \bm{\theta}_{\text{prop}})}{\pi(\bm{\theta}_{\text{cur}})g(\bm{\theta}_{\text{prop}} \mid \bm{\theta}_{\text{cur}})}\right) + L_2.
    \label{MBaccept}
\end{align}
Notably, the above minibatch acceptance rule is only accurate when, approximately, $(n^2/B)\sqrt{(n - B)/(n - 1)}\sigma_{\Lambda}^2 \leq 3$.  Thus, when implementing this minibatch algorithm in practice, we choose a cutoff $c \leq 3$ and then sample a sufficient minibatch size to ensure $(n^2/B)\sqrt{(n - B)/(n - 1)}\sigma_{\Lambda}^2 \leq c$.  To do so, we first obtain an estimate of $\sigma^2_\Lambda$ from an initial batch of $\Lambda_i$ then increase this initial batch size to ensure the condition $(n^2/B)\sqrt{(n - B)/(n - 1)}\sigma_{\Lambda}^2 \leq c$ is met so the minibatch acceptance rule in \eqref{MBaccept} can be used.

\subsection{Implementation Details}\label{implement}
Algorithms \ref{alg:miniBarker} and \ref{alg:miniMH} outline two possible minibatch MCMC algorithms for fitting GPs to data.  Algorithm \ref{alg:miniBarker} uses the Barker accept-reject rule while Algorithm \ref{alg:miniMH} uses a minibatch approximation of the Metropolis-Hastings rule.  As there are various subtleties associated with each of these algorithms, we discuss the details of implementing these algorithms and some of their differences here.

First, given the results in \eqref{qDist} and \eqref{lbarDist}, the minibatch size required will need to increase with the total sample size $n$.  This is to be expected in that larger minibatches will be required to sufficiently approximate the true acceptance rule or quantities in the complete conditional distributions.  

Second, for the minibatch algorithm based on the Barker acceptance test in Algorithm \ref{alg:miniBarker}, the required minibatch size ($B$) will  depend on the proposal $\bm{\theta}_{\text{prop}}$.  This is because if $\|\bm{\theta}_{\text{prop}} - \bm{\theta}_{\text{cur}}\|$ is large then $\sigma^2_\Lambda$ will likely also be large due to large changes in the likelihood ratios $\Lambda_i$.  In our observation, this fact means that the minibatch acceptance rule will often result in slower mixing of the Markov chain due to the need to take smaller steps at each iteration.  Hence, our minibatch acceptance rule decreases the computation time per iteration but we have found that minibatch chains need to be run longer to achieve convergence.  This is consistent with the findings of \citet{johndrow2020no} in that there is no free lunch with minibatch MCMC algorithms.

Third, again for Algorithm \ref{alg:miniBarker}, we draw a Gaussian random variable $L_1^\star \sim \mathcal{N}(0, c-(n^2/B_{\bm{\theta}})\sqrt{(n - B_{\bm{\theta}})/(n - 1)}\sigma_{\Lambda}^2)$ which is added to the acceptance rule \eqref{MBaccept}.  This is done to avoid the computational expense of calculating the correction distribution $h(\cdot)$ at each iteration.  That is, note that in Algorithm \ref{alg:miniBarker} we estimate the correction distribution (see Figure \ref{fig:cdist}) for a fixed $c \leq 3$ outside of the for-loop. Each iteration of Algorithm \ref{alg:miniBarker}, however, will choose a batch size ($B_\theta$) so that $(n^2/B_{\bm{\theta}})\sqrt{(n - B_{\bm{\theta}})/(n - 1)}\sigma_{\Lambda}^2 \leq c$.  In the event that $(n^2/B_{\bm{\theta}})\sqrt{(n - B_{\bm{\theta}})/(n - 1)}\sigma_{\Lambda}^2 < c$, we also sample $L_1^\star$ to ensure a match to the pre-calculated correction distribution $L_2 \sim h(x \mid c)$.

Fourth, note that Algorithm \ref{alg:miniMH}, is setup to do $E$ epochs over the $M$ minibatches.  This can be done in Algorithm \ref{alg:miniMH} because it uses a fixed batch size to estimate the Metropolis-Hastings rule rather than adapt the batch size to approximate the Barker acceptance test.  This has a few advantages over Algorithm \ref{alg:miniBarker}.  The foremost advantage of this type of looping is that the data is split into minibatches once rather than a different random sample taken at each iteration which can be time consuming for very large datasets.  Notably, the split into minibatches could occur after each epoch but this will add to computation time.  A second advantage of this setup is that it ensures that each data point is used to update parameters.  Under Algorithm \ref{alg:miniBarker} note that some data points may never be used by random chance.

Finally, the choice of batch size in both algorithms will influence the approximation.  Notably, as we show below, the approximation to the full posterior will improve as the batch size increases.  This makes the choice of batch size fundamentally different than the batch size used in, say, stochastic gradient descent which speeds up convergence by not as easily getting stuck in local modes.  In our case of using minibatching for posterior sampling, we want to use as large of a batch size as can be handled computationally.

\begin{algorithm}
\caption{A Minibatch MCMC Algorithm for GPs using Barker's Test}\label{alg:miniBarker}
\begin{algorithmic}[1]
\State Initialize $\beta_{\text{cur}}$, $\sigma^2_{\text{cur}}$, $\bm{\theta}_{\text{cur}}$
\State Choose $B_\beta$ and $B_{\sigma^2}$ for minibatch approximations to complete conditionals
\State Set initial batch size $B_{\text{init}}$ and increments $B_{\text{inc}}$ for Barker test
\State Set logistic approximation cutoff $c \leq 3$
\State Estimate correction distribution $h(x \mid c)$ via penalized least squares with positivity constraints 
\For{$i$ in $1, \dots I$}
    \State
    \For{$p$ in $0, \dots P$}
        \State Sample minibatch $\mathcal{B}_{\beta}$ of size $B_\beta$ from $\{1, \dots, n\}$
        \State Draw $\beta_{p}$ from \eqref{betaCC} using the minibatch approximation in \eqref{miniCC}
    \EndFor
    \State 
    \State Sample minibatch $\mathcal{B}_{\sigma^2}$ of size $B_{\sigma^2}$ from $\{1, \dots, n\}$
    \State Draw $\sigma^2$ from \eqref{sig2CC} using the minibatch approximation in \eqref{miniCC}
    \State 
    \State Propose $\bm{\theta}_{\text{prop}} \sim g()$
    \State Sample $\mathcal{B}_{\bm{\theta}}$ of size $B_{\text{init}}$ from $\{1, \dots, n\}$ 
    \State Set $\sigma^2_{\Lambda} = \mathbb{V}\text{ar}(\{\Lambda_i\}_{i \in \mathcal{B}_{\bm{\theta}}})$
    \While{$(n^2/B_{\bm{\theta}})\sqrt{(n - B_{\bm{\theta}})/(n - 1)}\sigma_{\Lambda}^2 > c$}
        \State Append $\mathcal{B}_{\bm{\theta}}$ by sampling an additional $B_{\text{inc}}$ from $\{1,\dots,n\} \setminus \mathcal{B}_{\bm{\theta}}$
        \State Set $B_\theta \leftarrow B_\theta+B_{\text{inc}}$
        \State Recalculate $\sigma^2_{\Lambda} = \text{Var}(\{\Lambda_i\}_{i \in \mathcal{B}_{\bm{\theta}}})$
    \EndWhile
    \State Draw $L_2 \sim h(\cdot \mid c)$
    \State Draw $L_1^\star \sim \mathcal{N}(0, c-(n^2/B_{\bm{\theta}})\sqrt{(n - B_{\bm{\theta}})/(n - 1)}\sigma_{\Lambda}^2)$
    \State Calculate $$\Delta(\bm{\theta}_{\text{prop}}, \bm{\theta}) = n\widebar{\Lambda}_B + \log\left(\frac{\pi(\bm{\theta}_{\text{prop}})g(\bm{\theta} \mid \bm{\theta}_{\text{prop}})}{\pi(\bm{\theta})g(\bm{\theta}_{\text{prop}} \mid \bm{\theta})}\right) + L_1^\star + L_2 $$
    \State $\bm{\theta} \gets \bm{\theta}_{\text{prop}}$ if $\Delta(\bm{\theta}_{\text{prop}}, \bm{\theta}) > 0$ using \eqref{MBaccept}
    \State
    \State Store $\bm{\beta}$, $\sigma^2$, $\bm{\theta}$
    \EndFor
\end{algorithmic}
\end{algorithm}

\begin{algorithm}
\caption{A Minibatch MCMC Algorithm for GPs using Metropolis-Hastings}\label{alg:miniMH}
\begin{algorithmic}[1]
\State Initialize $\beta_{\text{cur}}$, $\sigma^2_{\text{cur}}$, $\bm{\theta}_{\text{cur}}$
\State Set batch size $B$
\State Randomly divide the $n$ datapoints into $H$ batches of size $B$
\State \For{epoch in $1, \dots E$}
    \State \For{$h$ in $1,\dots,H$}
    
    \State \For{$p$ in $0, \dots P$}
        \State Draw $\beta_{p}$ from \eqref{betaCC} using minibatch $h$ in \eqref{miniCC}
    \EndFor
    \State 
    \State Draw $\sigma^2$ from \eqref{sig2CC} using minibatch $h$ in \eqref{miniCC}
    \State 
    \State Propose $\bm{\theta}_{\text{prop}} \sim g()$
    \State Let $L = -\log(U)$ where $U \sim \mathcal{U}(0,1)$ and using minibatch $h$ calculate $$\Delta(\bm{\theta}_{\text{prop}}, \bm{\theta}_{\text{cur}}) = n\widebar{\Lambda}_B + \log\left(\frac{\pi(\bm{\theta}_{\text{prop}})g(\bm{\theta} \mid \bm{\theta}_{\text{prop}})}{\pi(\bm{\theta}_{\text{cur}})g(\bm{\theta}_{\text{prop}} \mid \bm{\theta})}\right) + L $$
    \State $\bm{\theta} \gets \bm{\theta}_{\text{prop}}$ if $\Delta(\bm{\theta}_{\text{prop}}, \bm{\theta}) > 0$ using \eqref{MBaccept}
    \State Store $\bm{\beta}$, $\sigma^2$, $\bm{\theta}$
    \State
    \EndFor
    \EndFor
\end{algorithmic}
\end{algorithm}


\section{Examples}\label{examples}

\subsection{Small Data Simulation Study}\label{simStudies}

To explore the various algorithms explained above, we carry out a simulation study using 50 simulated data sets with $n=8000$ and the locations $\bm{s}_i$ simulated uniformly on the unit square.  Notably this is not, by any means, considered ``big data'' but we seek to evaluate the effectiveness of our minibatch algorithms relative to the full Gaussian process model.  Hence, for this simulation study, we let $n=8000$ because the full model in \eqref{fullLike} can be used. Of the $n$ data points, 1600 were randomly set aside as a test set, leaving 6400 to be used for model fitting. We simulate data from \eqref{fullModel} with covariance given by \eqref{covFX} using a stationary exponential correlation function with range $\phi$ and nugget $\omega$.  We set $\bm{\beta} = (0,1,-5)'$, $\sigma^2 = 1$, $\omega = 0.5$, and $\phi = 0.236$ which corresponds to an effective spatial range of $\sqrt{2}$ (half the maximum possible distance on the unit square).

To allow for accurate comparison between the algorithms, we used the same prior values across all algorithms. Specifically, we assume $\bm{\beta} \sim N(0, 1000\bm{I})$, $\sigma^2 \sim \mathcal{IG}( 0.01, 0.01)$ where $\mathcal{IG}(a,b)$ is the inverse gamma distribution with shape $a$ and rate $b$.  The parameters for the correlation function, $\omega$ and $\phi$, are notoriously difficult to estimate so, generally, bounded or discrete priors are commonly used \citep[see][]{zhang2004inconsistent,kaufman2013role, saha2023incorporating}.  Because a discrete prior will result in a full Gibbs algorithm, we consider both types of priors.  First, for continuous priors, we assume that $\omega \in [0,1]$ and $\phi \in [\phi_{\min}, \phi_{\max}]$.  But, in order to better propose values for $\omega$ and $\phi$, we transformed these to the real line via,
\begin{align}
    \phi^\star &= \log\left( \frac{(\phi-\phi_{min})/(\phi_{max}-\phi_{\min})}{1-((\phi-\phi_{\min})/(\phi_{\max}-\phi_{\min}))} \right) \\
    \omega^\star &= \log\left( \frac{\omega}{1-\omega} \right)
\end{align}
and assume $\omega^\star \sim N(0, 3)$, $\phi \sim N(0, 3)$.  The mean of zero means that, a priori, we expect that these parameters will be centered at the midpoint between their max and min values and a variance of 3 results in high uncertainty.  Finally, for a discrete prior, we choose 20 values for $\omega$ and $\phi$ between $[0,1]$ and $[\phi_{\min}, \phi_{\max}]$, respectively, and use a discrete uniform prior.

To evaluate and compare various algorithms with the proposed minibatch algorithms, we ran each of the following for 12,800 iterations and discarded the first 6,400 iterations as burn in (evaluations of trace plots suggested this was sufficient for convergence): 
\begin{itemize}
\item (Full) A Metropolis-within-Gibbs algorithm on the full model in Equation \eqref{fullLike} using the complete conditional distributions and Metropolis acceptance probability discussed in Section \ref{prelim};
\item (NN) The same algorithm as Full except we use the nearest neighbor (Vecchia) approximation in the complete conditionals and Metropolis acceptance probability;
\item (Barker) Algorithm 1;
\item (FB) Algorithm 2 where we use a specified fixed percentage of the full data as the minibatch size (i.e.\ a \textit{f}ixed \textit{b}atch size).
\end{itemize}
For Algorithm 2, we split the data into 2, 4, 8, 16, 32, and 64 smaller minibatches equating to batch sizes of 50\%, 25\%, 12.5\%, 6.25\%, 3.125\%, and 1.5625\% of the available data. This variation in batch size will allow us to examine how the amount of data in each minibatch affects the quality of the posterior approximations. 

\begin{figure}[tb]
    \centering
    \includegraphics[scale=.475]{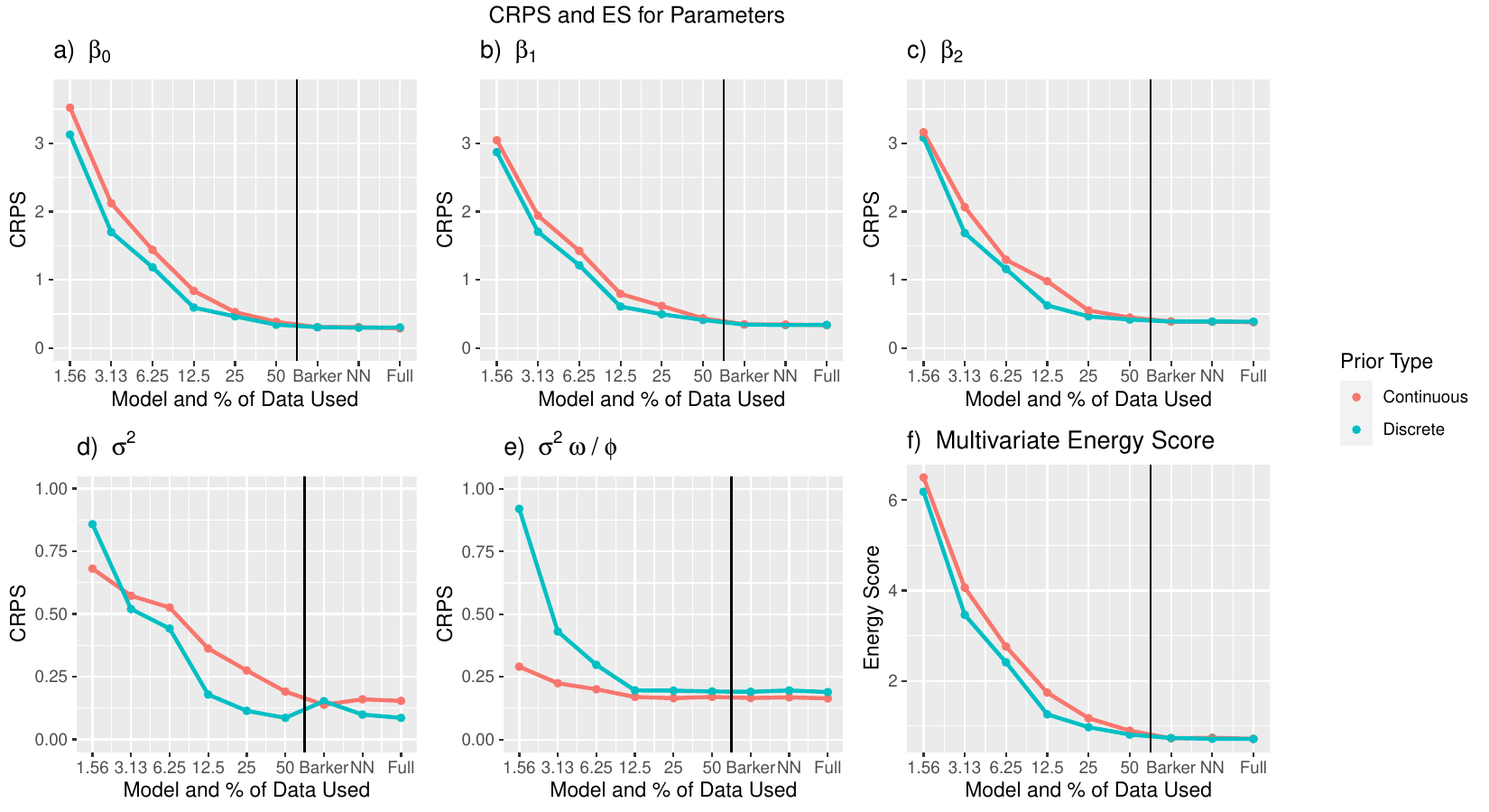}
    \caption{CRPS values for the different algorithms across different parameters.  Data points to the left of the vertical line corresponds to Algorithm 2 with the given percentage of the full data used as the minibatch size.  The different colors correspond to the use of a continuous or discrete prior for $\phi$ and $\omega$.}
    \label{fig:cdistsmall}
\end{figure}

Figure \ref{fig:cdistsmall} displays the continuous rank probability score (CRPS; \citealt{gneiting2007strictly}) achieved by the various algorithms for the different parameters of the Gaussian process model.  We note that the parameter $\sigma^2 \omega/\phi$ is mentioned by \citet{zhang2004inconsistent} as the parameter that is able to be consistently estimated from a single realization of a Gaussian process so we include it here for comparison.  From Figure \ref{fig:cdistsmall}, the Barker, NN and Full algorithms all achieve low CRPS values indicating posterior distributions that capture the true parameter value. By comparison, the FB algorithms show increasing ability to estimate parameters (as indicated by decreasing CRPS) as the minibatch size increases (note that 50\% corresponds to a minibatch size of 50\% of $n$ or $0.5\times6400 = 3200$).  Finally, this same pattern persists between the discrete and continuous priors for $\phi$ and $\omega$ except that the discrete priors tended to have slightly lower CRPS values. 

As a last comparison, we compare each algorithm in terms of the multivariate energy score (MES; \citealt{gneiting2008assessing}) to evaluate the overall accuracy of parameter estimates. The bottom right panel of Figure \ref{fig:cdist} shows that MES values for each algorithm, averaged across all data sets, revealing a similar pattern to the CRPS values for individual parameters.

\begin{figure}[tb]
    \centering
    \includegraphics[scale=.575]{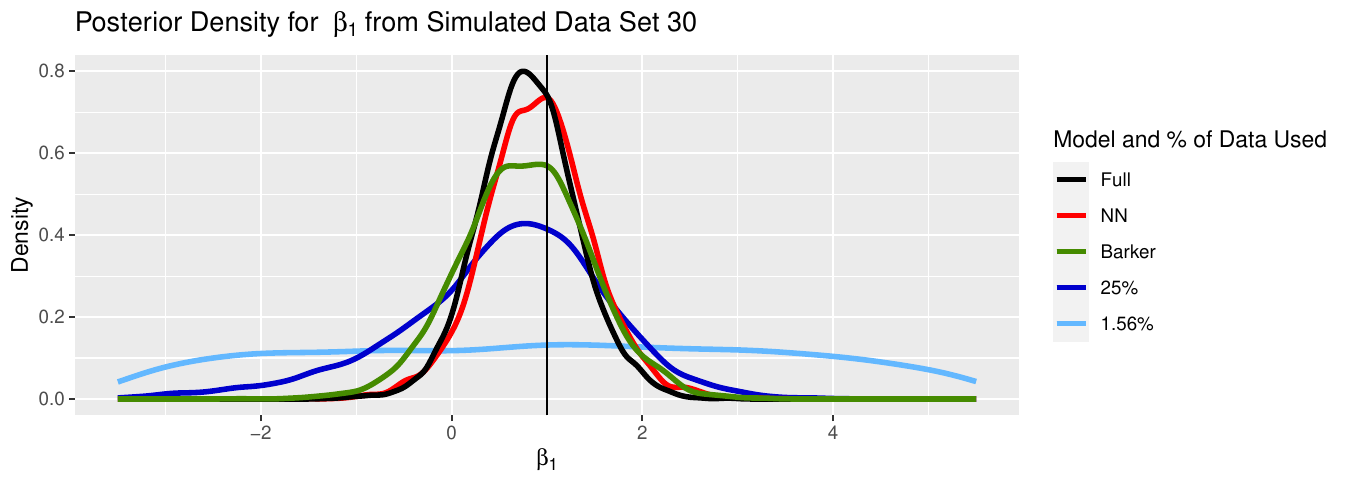}
    \caption{Posterior density of Beta1 for the 30th simulated data set.}
    \label{fig:postDists}
\end{figure}

To further elucidate the impact of the minibatch size on the posterior distribution, Figure \ref{fig:postDists} displays density plots for the $\beta$ parameters.  Note that the posterior distribution for each algorithm is centered at the true value but the decrease in CRPS observed in Figure \ref{fig:cdistsmall} can be attributed to an increase in the variance of the posterior distribution.  That is, algorithms based on minibatching have increased posterior variance relative to posterior distributions using the full data.







\begin{figure}[tb]
    \centering
    \includegraphics[scale=.575]{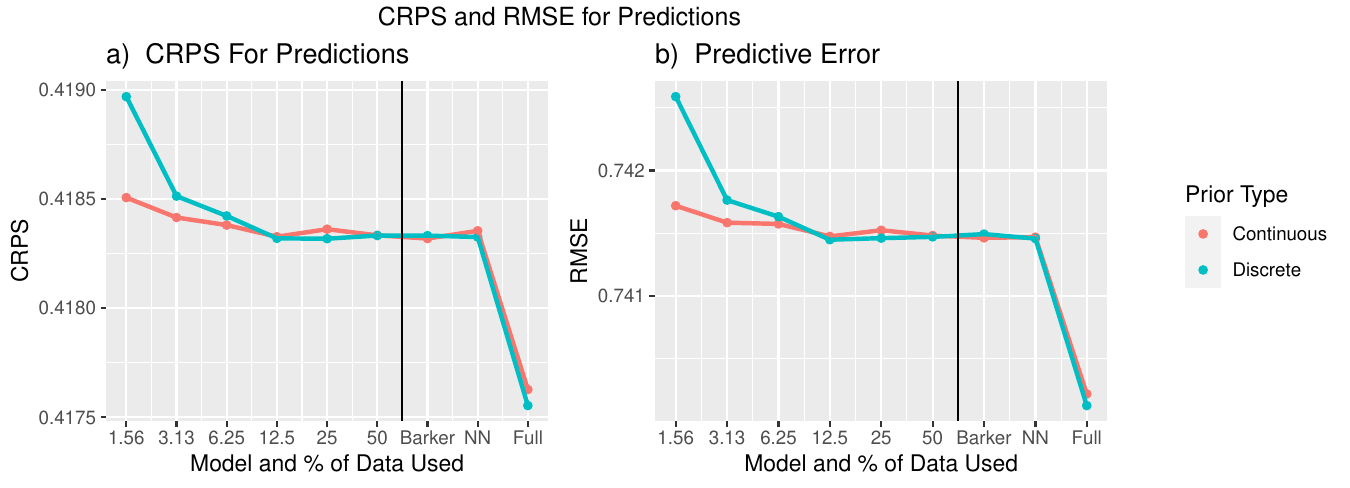}
    \caption{Predictive accuracy for small study.}
    \label{fig:predRMSE}
\end{figure}

The predictive accuracy of predictions generated under each of the above algorithms are shown in Figure \ref{fig:predRMSE} in terms of root mean square error (RMSE) and CRPS.  Importantly, Figure \ref{fig:predRMSE} shows that RMSE and CRPS values are effectively equivalent for all algorithms.  The result that the predictions are equivalent under minibatching relative to the full data is substantial and aligns with the results in \citet{saha2023incorporating}. This shows that minibatching offers computational advantages with no apparent lack of predictive ability.

The motivation behind both minibatch algorithms is computational savings. To this end, we found that as the amount of data included in each minibatch decreases, so does the computation time. Using the full model with either continuous or discrete priors on the spatial parameters took about 4 times longer than the nearest neighbor approximation. However, using half of the data in each minibatch (FB2) takes about half of the time that the nearest neighbor algorithm takes. Using 16 minibatches (all of equal size with about 6\% of the data) takes only 10\% of the time of the nearest neighbor algorithm.  The Barker algorithms chose minibatch sizes of about 25\% of the data and, therefore, had similar computation to the FB4 algorithm.

\subsection{Large Data Simulation Studies}\label{LargesimStudies}

In a larger scale study, we also carried out a simulation study using the same setup that was used in Section \ref{simStudies} above but with $n=120,000$.  We fit all the same algorithms at the same settings used in the previous simulation study but we omit the full model because it is not reasonable to use on this size of dataset.

The full results with figures similar to that shown above are given in the supplementary material but we offer a summary of the findings here.  First, increasing batch size resulted in increasing performance in terms of CRPS.  Specifically, a 50\% minibatch size was nearly indistinguishable from the nearest neighbor model with all of the data.  While a 25\% minibatch size had strong results, decreasing the minibatch size further often resulted in unacceptable performance.  The Barker algorithm (Algorithm \ref{barker}) would often result in a minibatch size that was too small for acceptable CRPS performance relative to the nearest neighbor algorithm.

In terms of predictive accuracy, we again saw that the minibatch size had no effect on the predictive performance.  Any of the minibatch algorithms were effectively equal in predictive accuracy to the nearest neighbor model.

Finally, because the minibatch sizes were chosen as a percentage of the data, the computation times relative to the nearest neighbor model stayed about the same as in the smaller data simulation study in Section \ref{simStudies}.  For example, a 50\% minibatch size takes about half the time as the nearest neighbor model with the full data but computation time decreased with minibatch size.

\subsection{Real Data Applications}\label{RealStudies}

\begin{figure}[tb]
    \centering
    \includegraphics[scale=.575]{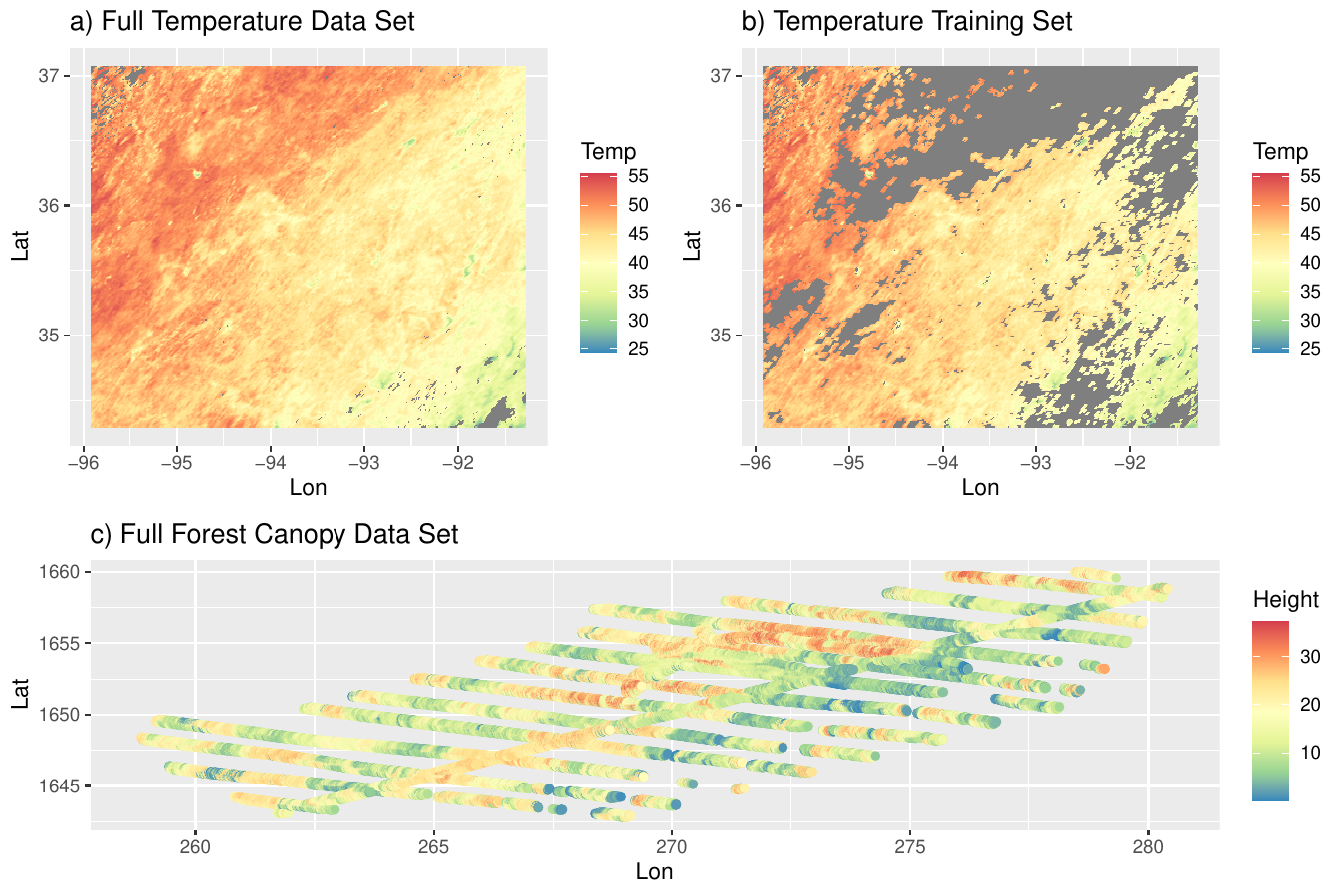}
    \caption{The satellite data (top row) and the forest canopy height data (bottom row).}
    \label{fig:realdata}
\end{figure}

In this section we consider applying the minibatch algorithms to two different real datasets.  First, we use the real satellite observations from \citet{heaton2019case} to compare the approach used here with those of other models.  Second, we apply our minibatch methods to the forest canopy height (FCH) dataset available in the \texttt{spNNGP} \texttt{R} library \citep{spNNGP}.  Because our focus is on performance of the minibatch algorithms, we refer to the given citations for the scientific details for both of these datasets but Figure \ref{fig:realdata} displays the datasets.

For both of these datasets, we applied the same minibatch algorithms as were used for the simulation studies above.  That is, we fit the model using Algorithm \ref{alg:miniBarker} and Algorithm \ref{alg:miniMH} at various minibatch sizes, along with the nearest neighbor algorithm to serve as a benchmark.   Further, we fit every model twice, once with a continuous prior for $\phi$ and $\omega$ and once with a discrete prior. We measure the computation time associated with each algorithm and along with multiple metrics to compare the effectiveness and accuracy of each approach. 

In the previous simulation studies, accuracy of the posterior distribution of model parameters was assessed via CRPS because the true parameter values were known. However, for real data applications, the true parameter values are unknown. Hence, here we compare posterior summaries under the various algorithms.  Table \ref{con_value} displays the posterior mean, standard deviation and width of a 95\% credible interval for the various algorithms for the parameter $\sigma^2 \omega/\phi$ which again, according to \citet{zhang2004inconsistent}, is the identifiable parameter in a Gaussian process model.  First, from Table \ref{con_value} note that the estimates of the posterior means are consistent across algorithms while the posterior standard deviations increase as the minibatch size shrinks.  This result further confirms the results from the simulation studies along with those in \citet{wu2022mini} that minibatching results in a tempered posterior distribution with smaller minibatches corresponding to a higher temperature.  We note that Algorithm \ref{alg:miniBarker} (Barker) was selecting a minibatch size of, approximately, 8000 which corresponds to about 8\% of the data as a minibatch and thus had a higher posterior variance.

\begin{table}[tb]
\centering
\begin{tabular}{c|ccc|ccc}
  \multicolumn{1}{c}{} & \multicolumn{3}{|c|}{Satellite Data} & \multicolumn{3}{c}{Forest Data} \\
  \hline
  Algorithm & Mean & SD & Width & Mean & SD & Width \\ 
  \hline
  NN & 45.73 & 0.24 & 0.87 & 353.48 & 4.45 & 15.41 \\ 
  BG & 43.89 & 2.55 & 9.28 & 327.50 & 17.62 & 64.35 \\ 
  FB2 & 45.73 & 0.64 & 1.88 & 353.51 & 7.20 & 25.51 \\ 
  FB4 & 45.71 & 0.88 & 2.80 & 351.85 & 8.36 & 29.65 \\ 
  FB8 & 45.71 & 1.07 & 3.73 & 350.84 & 10.65 & 37.51 \\ 
  FB16 & 45.77 & 1.41 & 5.00 & 351.67 & 13.39 & 48.50 \\ 
  FB32 & 45.67 & 1.89 & 7.42 & 349.57 & 19.07 & 69.37 \\ 
  FB64 & 45.59 & 3.03 & 11.37 & 350.15 & 27.81 & 97.65 \\ 
   \hline
\end{tabular}
\caption{Posterior Estimates of $\sigma^2 \omega / \phi$ using a continuous prior and both application data sets.  Note the increase in credible interval width as the minibatch size decreases.}
\label{con_value}
\end{table}

\begin{figure}[tb]
    \centering
    \includegraphics[scale=.575]{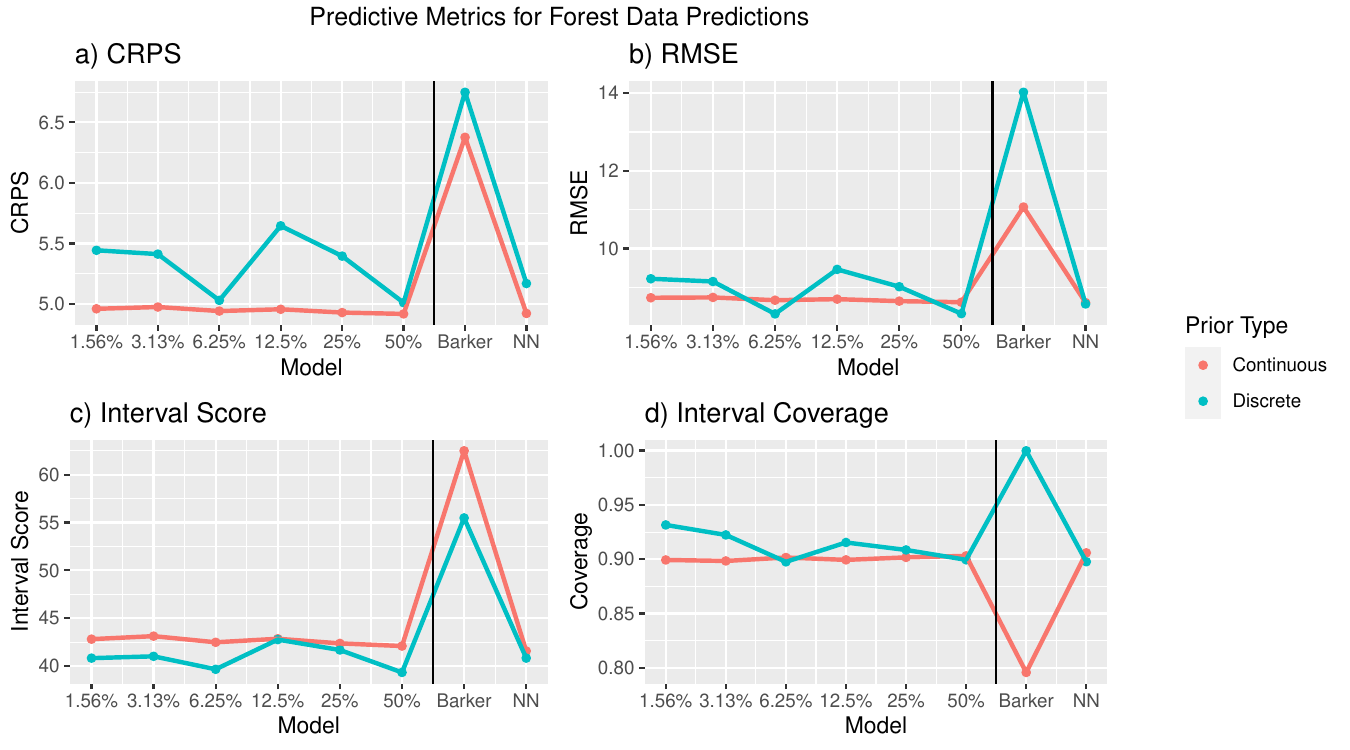}
    \caption{Predictive Metrics for Forest Data Predictions}
    \label{fig:ForestPreds}
\end{figure}

To measure predictive accuracy for the real data, we split the data into training and test sets.  For both the satellite and forest data, we used the train-test split provided in the datasets by the original users.  Table \ref{Discrete_Satellite_Preds} displays the same predictive diagnostics for the satellite data as was calculated in \citet{heaton2019case} while Figure \ref{fig:ForestPreds} displays predictive diagnostics for the forest data.  For the satellite data, the predictive diagnostics under the minibatch algorithms, while not the best as has been applied to this data, are comparable.  For the forest data, each of the fixed batch algorithms, regardless of minibatch size, achieved predictive performance comparable to that of the nearest neighbor model.  The Barker algorithm, however, had worse predictive accuracy.
\begin{table}[tb]
\centering
\label{Discrete_Satellite_Preds}
\begin{tabular}{ccccccc}
  \hline
Method & MAE & RPMSE & CRPS & INT & WID & CVG \\ 
  \hline
NN & 1.38 & 1.86 & 0.99 & 0.23 & 4.89 & 0.87 \\ 
  BG & 1.36 & 1.79 & 0.94 & 0.19 & 5.50 & 0.93 \\ 
  FB2 & 1.63 & 2.14 & 1.22 & 0.39 & 4.37 & 0.74 \\ 
  FB4 & 1.46 & 1.95 & 1.06 & 0.27 & 4.70 & 0.83 \\ 
  FB8 & 1.47 & 1.97 & 1.07 & 0.28 & 4.71 & 0.83 \\ 
  FB16 & 1.62 & 2.13 & 1.21 & 0.38 & 4.37 & 0.75 \\ 
  FB32 & 1.62 & 2.12 & 1.21 & 0.38 & 4.39 & 0.75 \\ 
  FB64 & 1.46 & 1.95 & 1.05 & 0.25 & 4.82 & 0.84 \\ 
   \hline
\end{tabular}
\caption{Predictive accuracy for discrete models on the satellite data.  These are comparable to the competition numbers shown in \citet{heaton2019case}.}
\end{table}

\section{Conclusion and Further Research}\label{conc}

In this research, we presented possible approaches for using minibatching with MCMC algorithms when fitting Gaussian processes to large spatial datasets.  In terms of parameter estimation, minibatch sizes of greater than about 25\% of the data resulted in comparable estimation performance to algorithms that used all of the data.  While small minibatches resulted in poor estimation of parameters, such minibatches seemed to have no impact on the predictive performance.

Generally, we presented two algorithms for using minibatching within MCMC algorithms: one based on an adaptive minibatch size (Barker) and one based on a fixed minibatch size.  The advantage of the Barker approach is that the minibatch size needed to achieve a sufficient approximation of the acceptance rule is automatically chosen within the algorithm.  In our studies, however, this advantage was offset by the computational time needed to find the appropriate batch size.  Hence, based on our experience, we recommend the fixed batch approach as it it faster and still gives good posterior performance.

In practice, the batch size $B$ will be chosen based on the computational demand and in this work we evaluated the performance based on various fixed batch sizes.  However, we note that the result in \eqref{qDist} can guide the choice of minibatch size in a few different ways.  First, given an approximation of, say, $\sigma^2_{q_j}$ (perhaps obtained using the starting values of the MCMC algorithm or from the first several draws of the algorithm), $B$ can be chosen so that $\mathbb{V}\text{ar}(n\widebar{q}_{j,B})$ in \eqref{qDist} is less than a certain threshold.  This will guarantee that the minibatch meets certain approximation criteria. Alternatively, at any given iteration of the MCMC algorithm, an estimate of $\sigma^2_{q_j}$ can be obtained based on an initial batch size (similar to Algorithm \ref{barker}) and then $B$ can be chosen to meet an approximation criteria resulting in a different batch size at each iteration.  However, varying the batch size by iteration can slow computation.

We note that this research focused on stationary GPs because stationarity is commonly assumed in spatial analyses.  However, minibatching might impact anisotropy, nonstationarity or spatio-temporal GPs differently.  For example, nonstationary models require local information because the correlation changes over the spatial domain.  As such, minibatching might result in a loss of local information leading to more variability in the posterior.  Future work needs to focus on the use of these methods on more general correlation structures.

Our approach approximates accept-reject rules based on minibatch samples.  As such, these approximations can be used for any spatial model - not just the one detailed in Equation \eqref{fullModel} here. For example, similar approaches to minibatching could be used to fit non-Gaussian spatial linear models.  We plan to investigate this in future work.


\bibliographystyle{ba}
\bibliography{MBMCMC}

\begin{thebibliography}{39}
\newcommand{\enquote}[1]{``#1''}
\expandafter\ifx\csname natexlab\endcsname\relax\def\natexlab#1{#1}\fi
\expandafter\ifx\csname url\endcsname\relax
  \def\url#1{{\tt #1}}\fi
\expandafter\ifx\csname urlprefix\endcsname\relax\def\urlprefix{URL }\fi
\ifx\endbibitem\undefined \let\endbibitem\relax\fi

\bibitem[{Abdulah et~al.(2022)Abdulah, Alamri, Nag, Sun, Ltaief, Keyes, and
  Genton}]{abdulah2022second}
Abdulah, S., Alamri, F., Nag, P., Sun, Y., Ltaief, H., Keyes, D.~E., and
  Genton, M.~G. (2022).
\newblock \enquote{The Second Competition on Spatial Statistics for Large
  Datasets.}
\newblock {\em arXiv preprint arXiv:2211.03119\/}.
\endbibitem

\bibitem[{Abdulah et~al.(2018)Abdulah, Ltaief, Sun, Genton, and
  Keyes}]{abdulah2018exageostat}
Abdulah, S., Ltaief, H., Sun, Y., Genton, M.~G., and Keyes, D.~E. (2018).
\newblock \enquote{Exageostat: A high performance unified software for
  geostatistics on manycore systems.}
\newblock {\em IEEE Transactions on Parallel and Distributed Systems\/},
  29(12): 2771--2784.
\endbibitem

\bibitem[{Banerjee et~al.(2003)Banerjee, Carlin, and
  Gelfand}]{banerjee2003hierarchical}
Banerjee, S., Carlin, B.~P., and Gelfand, A.~E. (2003).
\newblock {\em Hierarchical modeling and analysis for spatial data\/}.
\newblock Chapman and Hall/CRC.
\endbibitem

\bibitem[{Banerjee et~al.(2008)Banerjee, Gelfand, Finley, and
  Sang}]{banerjee2008gaussian}
Banerjee, S., Gelfand, A.~E., Finley, A.~O., and Sang, H. (2008).
\newblock \enquote{Gaussian predictive process models for large spatial data
  sets.}
\newblock {\em Journal of the Royal Statistical Society: Series B (Statistical
  Methodology)\/}, 70(4): 825--848.
\endbibitem

\bibitem[{Bardenet et~al.(2014)Bardenet, Doucet, and
  Holmes}]{bardenet2014towards}
Bardenet, R., Doucet, A., and Holmes, C. (2014).
\newblock \enquote{Towards scaling up Markov chain Monte Carlo: an adaptive
  subsampling approach.}
\newblock In {\em International conference on machine learning\/}, 405--413.
  PMLR.
\endbibitem

\bibitem[{Bardenet et~al.(2017)Bardenet, Doucet, and
  Holmes}]{bardenet2017markov}
Bardenet, R., Doucet, A., and Holmes, C.~C. (2017).
\newblock \enquote{On Markov chain Monte Carlo methods for tall data.}
\newblock {\em Journal of Machine Learning Research\/}, 18(47).
\endbibitem

\bibitem[{Barker(1965)}]{barker1965monte}
Barker, A.~A. (1965).
\newblock \enquote{Monte carlo calculations of the radial distribution
  functions for a proton? electron plasma.}
\newblock {\em Australian Journal of Physics\/}, 18(2): 119--134.
\endbibitem

\bibitem[{Bottou(2012)}]{bottou2012stochastic}
Bottou, L. (2012).
\newblock \enquote{Stochastic gradient descent tricks.}
\newblock In {\em Neural networks: Tricks of the trade\/}, 421--436. Springer.
\endbibitem

\bibitem[{Cressie(2015)}]{cressie2015statistics}
Cressie, N. (2015).
\newblock {\em Statistics for spatial data\/}.
\newblock John Wiley \& Sons.
\endbibitem

\bibitem[{Cressie and Johannesson(2008)}]{cressie2008fixed}
Cressie, N. and Johannesson, G. (2008).
\newblock \enquote{Fixed rank kriging for very large spatial data sets.}
\newblock {\em Journal of the Royal Statistical Society: Series B (Statistical
  Methodology)\/}, 70(1): 209--226.
\endbibitem

\bibitem[{Datta et~al.(2016{\natexlab{a}})Datta, Banerjee, Finley, and
  Gelfand}]{datta2016hierarchical}
Datta, A., Banerjee, S., Finley, A.~O., and Gelfand, A.~E.
  (2016{\natexlab{a}}).
\newblock \enquote{Hierarchical nearest-neighbor {G}aussian process models for
  large geostatistical datasets.}
\newblock {\em Journal of the American Statistical Association\/}, 111(514):
  800--812.
\endbibitem

\bibitem[{Datta et~al.(2016{\natexlab{b}})Datta, Banerjee, Finley, and
  Gelfand}]{datta2016cholesky}
--- (2016{\natexlab{b}}).
\newblock \enquote{On nearest-neighbor {G}aussian process models for massive
  spatial data.}
\newblock {\em Wiley Interdisciplinary Reviews: Computational Statistics\/},
  8(5): 162--171.
\newline\urlprefix\url{http://dx.doi.org/10.1002/wics.1383}
\endbibitem

\bibitem[{De~Sa et~al.(2018)De~Sa, Chen, and Wong}]{de2018minibatch}
De~Sa, C., Chen, V., and Wong, W. (2018).
\newblock \enquote{Minibatch gibbs sampling on large graphical models.}
\newblock In {\em International Conference on Machine Learning\/}, 1173--1181.
\endbibitem

\bibitem[{Finley et~al.(2022)Finley, Datta, and Banerjee}]{spNNGP}
Finley, A.~O., Datta, A., and Banerjee, S. (2022).
\newblock \enquote{{spNNGP} {R} Package for Nearest Neighbor {G}aussian Process
  Models.}
\newblock {\em Journal of Statistical Software\/}, 103(5): 1--40.
\endbibitem

\bibitem[{Furrer et~al.(2006)Furrer, Genton, and Nychka}]{furrer2006covariance}
Furrer, R., Genton, M.~G., and Nychka, D. (2006).
\newblock \enquote{Covariance tapering for interpolation of large spatial
  datasets.}
\newblock {\em Journal of Computational and Graphical Statistics\/}, 15(3):
  502--523.
\endbibitem

\bibitem[{Gelfand et~al.(2010)Gelfand, Diggle, Guttorp, and
  Fuentes}]{gelfand2010handbook}
Gelfand, A.~E., Diggle, P., Guttorp, P., and Fuentes, M. (2010).
\newblock {\em Handbook of spatial statistics\/}.
\newblock CRC press.
\endbibitem

\bibitem[{Gelfand and Schliep(2016)}]{gelfand2016spatial}
Gelfand, A.~E. and Schliep, E.~M. (2016).
\newblock \enquote{Spatial statistics and Gaussian processes: A beautiful
  marriage.}
\newblock {\em Spatial Statistics\/}, 18: 86--104.
\endbibitem

\bibitem[{Genton(2001)}]{genton2001classes}
Genton, M.~G. (2001).
\newblock \enquote{Classes of kernels for machine learning: a statistics
  perspective.}
\newblock {\em Journal of machine learning research\/}, 2(Dec): 299--312.
\endbibitem

\bibitem[{Gneiting and Raftery(2007)}]{gneiting2007strictly}
Gneiting, T. and Raftery, A.~E. (2007).
\newblock \enquote{Strictly proper scoring rules, prediction, and estimation.}
\newblock {\em Journal of the American statistical Association\/}, 102(477):
  359--378.
\endbibitem

\bibitem[{Gneiting et~al.(2008)Gneiting, Stanberry, Grimit, Held, and
  Johnson}]{gneiting2008assessing}
Gneiting, T., Stanberry, L.~I., Grimit, E.~P., Held, L., and Johnson, N.~A.
  (2008).
\newblock \enquote{Assessing probabilistic forecasts of multivariate
  quantities, with an application to ensemble predictions of surface winds.}
\newblock {\em Test\/}, 17: 211--235.
\endbibitem

\bibitem[{Heaton et~al.(2017)Heaton, Christensen, and
  Terres}]{heaton2017nonstationary}
Heaton, M.~J., Christensen, W.~F., and Terres, M.~A. (2017).
\newblock \enquote{Nonstationary Gaussian process models using spatial
  hierarchical clustering from finite differences.}
\newblock {\em Technometrics\/}, 59(1): 93--101.
\endbibitem

\bibitem[{Heaton et~al.(2019)Heaton, Datta, Finley, Furrer, Guinness,
  Guhaniyogi, Gerber, Gramacy, Hammerling, Katzfuss et~al.}]{heaton2019case}
Heaton, M.~J., Datta, A., Finley, A.~O., Furrer, R., Guinness, J., Guhaniyogi,
  R., Gerber, F., Gramacy, R.~B., Hammerling, D., Katzfuss, M., et~al. (2019).
\newblock \enquote{A case study competition among methods for analyzing large
  spatial data.}
\newblock {\em Journal of Agricultural, Biological and Environmental
  Statistics\/}, 24(3): 398--425.
\endbibitem

\bibitem[{Huang et~al.(2021)Huang, Abdulah, Sun, Ltaief, Keyes, and
  Genton}]{huang2021competition}
Huang, H., Abdulah, S., Sun, Y., Ltaief, H., Keyes, D.~E., and Genton, M.~G.
  (2021).
\newblock \enquote{Competition on spatial statistics for large datasets.}
\newblock {\em Journal of Agricultural, Biological and Environmental
  Statistics\/}, 26(4): 580--595.
\endbibitem

\bibitem[{Johndrow et~al.(2020)Johndrow, Pillai, and Smith}]{johndrow2020no}
Johndrow, J.~E., Pillai, N.~S., and Smith, A. (2020).
\newblock \enquote{No free lunch for approximate MCMC.}
\newblock {\em arXiv preprint arXiv:2010.12514\/}.
\endbibitem

\bibitem[{Katzfuss and Guinness(2021)}]{katzfuss2021general}
Katzfuss, M. and Guinness, J. (2021).
\newblock \enquote{A general framework for {V}ecchia approximations of
  {G}aussian processes.}
\newblock {\em Statistical Science\/}, 36(1): 124--141.
\endbibitem

\bibitem[{Kaufman and Shaby(2013)}]{kaufman2013role}
Kaufman, C. and Shaby, B.~A. (2013).
\newblock \enquote{The role of the range parameter for estimation and
  prediction in geostatistics.}
\newblock {\em Biometrika\/}, 100(2): 473--484.
\endbibitem

\bibitem[{Kaufman et~al.(2008)Kaufman, Schervish, and
  Nychka}]{kaufman2008covariance}
Kaufman, C.~G., Schervish, M.~J., and Nychka, D.~W. (2008).
\newblock \enquote{Covariance tapering for likelihood-based estimation in large
  spatial data sets.}
\newblock {\em Journal of the American Statistical Association\/}, 103(484):
  1545--1555.
\endbibitem

\bibitem[{Konomi et~al.(2014)Konomi, Sang, and Mallick}]{konomi2014adaptive}
Konomi, B.~A., Sang, H., and Mallick, B.~K. (2014).
\newblock \enquote{Adaptive Bayesian nonstationary modeling for large spatial
  datasets using covariance approximations.}
\newblock {\em Journal of Computational and Graphical Statistics\/}, 23(3):
  802--829.
\endbibitem

\bibitem[{Korattikara et~al.(2014)Korattikara, Chen, and
  Welling}]{korattikara2014austerity}
Korattikara, A., Chen, Y., and Welling, M. (2014).
\newblock \enquote{Austerity in MCMC land: Cutting the Metropolis-Hastings
  budget.}
\newblock In {\em International Conference on Machine Learning\/}, 181--189.
  PMLR.
\endbibitem

\bibitem[{Li and Wong(2017)}]{li2017mini}
Li, D. and Wong, W.~H. (2017).
\newblock \enquote{Mini-batch tempered MCMC.}
\newblock {\em arXiv preprint arXiv:1707.09705\/}.
\endbibitem

\bibitem[{Mat{\'e}rn(1960)}]{matern1960spatial}
Mat{\'e}rn, B. (1960).
\newblock {\em Spatial variation\/}, volume~36.
\newblock Springer Science \& Business Media.
\endbibitem

\bibitem[{Noack et~al.(2023)Noack, Krishnan, Risser, and
  Reyes}]{noack2023exact}
Noack, M.~M., Krishnan, H., Risser, M.~D., and Reyes, K.~G. (2023).
\newblock \enquote{Exact Gaussian processes for massive datasets via
  non-stationary sparsity-discovering kernels.}
\newblock {\em Scientific reports\/}, 13(1): 3155.
\endbibitem

\bibitem[{Saha and Bradley(2023)}]{saha2023incorporating}
Saha, S. and Bradley, J.~R. (2023).
\newblock \enquote{Incorporating Subsampling into Bayesian Models for
  High-Dimensional Spatial Data.}
\newblock {\em arXiv preprint arXiv:2305.13221\/}.
\endbibitem

\bibitem[{Sang and Huang(2012)}]{sang2012full}
Sang, H. and Huang, J.~Z. (2012).
\newblock \enquote{A full scale approximation of covariance functions for large
  spatial data sets.}
\newblock {\em Journal of the Royal Statistical Society: Series B (Statistical
  Methodology)\/}, 74(1): 111--132.
\endbibitem

\bibitem[{Seita et~al.(2016)Seita, Pan, Chen, and Canny}]{seita2016efficient}
Seita, D., Pan, X., Chen, H., and Canny, J. (2016).
\newblock \enquote{An efficient minibatch acceptance test for
  metropolis-hastings.}
\newblock {\em arXiv preprint arXiv:1610.06848\/}.
\endbibitem

\bibitem[{Stein(2014)}]{stein2014limitations}
Stein, M.~L. (2014).
\newblock \enquote{Limitations on low rank approximations for covariance
  matrices of spatial data.}
\newblock {\em Spatial Statistics\/}, 8: 1--19.
\endbibitem

\bibitem[{Wikle et~al.(2019)Wikle, Zammit-Mangion, and
  Cressie}]{wikle2019spatio}
Wikle, C.~K., Zammit-Mangion, A., and Cressie, N. (2019).
\newblock {\em Spatio-temporal Statistics with R\/}.
\newblock Chapman and Hall/CRC.
\endbibitem

\bibitem[{Wu et~al.(2022)Wu, Rachel~Wang, and Wong}]{wu2022mini}
Wu, T.-Y., Rachel~Wang, Y., and Wong, W.~H. (2022).
\newblock \enquote{Mini-Batch Metropolis--Hastings With Reversible SGLD
  Proposal.}
\newblock {\em Journal of the American Statistical Association\/}, 117(537):
  386--394.
\endbibitem

\bibitem[{Zhang(2004)}]{zhang2004inconsistent}
Zhang, H. (2004).
\newblock \enquote{Inconsistent estimation and asymptotically equal
  interpolations in model-based geostatistics.}
\newblock {\em Journal of the American Statistical Association\/}, 99(465):
  250--261.
\endbibitem

\end{thebibliography}


\end{document}